\documentclass[a4paper,12pt]{article}
\usepackage{hyperref}
\usepackage{epsfig}
\usepackage{amssymb}
\usepackage{cancel}
\usepackage{setspace}
\usepackage[usenames,dvipsnames,svgnames,table]{xcolor}
\usepackage{amsmath}
\usepackage{graphicx}
\usepackage{subfigure}
\usepackage{url}
\usepackage{slashed}
\usepackage{booktabs}
\def\pd{\partial}

\newcommand{\be}{\begin{equation}}
\newcommand{\ee}{\end{equation}}
\newcommand{\br}{\begin{eqnarray}}
\newcommand{\bea}{\begin{eqnarray}}
\newcommand{\eea}{\end{eqnarray}}
\newcommand{\er}{\end{eqnarray}}
\newcommand{\ba}{\begin{array}}
\newcommand{\ea}{\end{array}}
\newcommand{\bi}{\begin{itemize}}
\newcommand{\ei}{\end{itemize}}
\newcommand{\bn}{\begin{enumerate}}
\newcommand{\en}{\end{enumerate}}
\newcommand{\bc}{\begin{center}}
\newcommand{\ec}{\end{center}}

\newcommand{\beq}{\begin{equation}}
\newcommand{\eeq}{\end{equation}}

\newcommand{\U}{\scriptscriptstyle U}
\newcommand{\D}{\scriptscriptstyle D}
\newcommand{\Q}{\scriptscriptstyle Q}

\newcommand{\E}{\scriptscriptstyle E}

\newcommand{\gsim}{\lower1.0ex\hbox{$\;\stackrel{\textstyle>}{\sim}\;$}}
\newcommand{\lsim}{\lower1.0ex\hbox{$\;\stackrel{\textstyle<}{\sim}\;$}}

\newcommand{\bs}{\begin{small}}
\newcommand{\es}{\end{small}}

\newcommand{\qui}{q_{{\scriptscriptstyle U}_{\!i}}}
\newcommand{\qdi}{q_{{\scriptscriptstyle D}_{\!i}}}

\newcommand{\LG}{\mathcal{L}}

\newcommand{\nn}{\nonumber} 
\newcommand{\dg}{\dagger}

\newcommand{\modu}[1]{\left|{#1}\right|}

\newcommand{\tw}{\theta_W}
\newcommand{\PL}{\operatorname{P_L}}
\newcommand{\PR}{\operatorname{P_R}}
\newcommand{\ON}[1]{\operatorname{#1}}

\newcommand{\g}{g_{\scriptscriptstyle{LR}}}

\begin{document}
\begin{center}
{\Large {\bf Radiative Yukawa Couplings in the Simplest
\\
\vspace*{0.3cm}
Left-Right Symmetric Model
}}\\
\vspace*{1.5cm}
{

 {\bf Emidio Gabrielli$^{a,b,c}$},   {\bf Luca Marzola$^{c,d}$},   {\bf Martti Raidal$^{c}$}
}\\

\vspace{0.5cm}
{\it
(a)  Dipartimento di Fisica, Theoretical section, Universit\`a di 
Trieste, \\ Strada Costiera 11, I-34151 Trieste, Italy \\ 
(b) INFN, Sezione di Trieste, Via Valerio 2, I-34127 Trieste, Italy
\\
(c) NICPB, R\"avala 10, Tallinn 10143, Estonia 
\\
(d) Institute of Physics, University of Tartu; W. Ostwaldi tn 1, 50411 Tartu, Estonia.  
\\[1mm] }

\vspace*{2cm}{\bf ABSTRACT}
\end{center}
\vspace{0.3cm}

\noindent
We revisit a recent solution to the flavour hierarchy problem based on the paradigm that Yukawa couplings are, rather than fundamental constants, effective low energy couplings radiatively generated by interactions in a hidden sector of the theory. In the present paper we show that the setup required by this scenario can be set by gauge invariance alone, provided that the Standard Model gauge group be extended to the left-right symmetric group of $SU(2)_L\times SU(2)_R\times U(1)_Y$. The simplest scheme in which Yukawa couplings are forbidden at the tree-level organises the right-handed fermions into doublets and presents an additional Higgs $SU(2)_R$ doublet, responsible for the spontaneous breaking of the $SU(2)_R$ gauge sector. The flavour and chiral symmetry breaking induced by the $SU(2)_R$ breaking is transferred at the one-loop level to the Standard Model via the dynamics of the hidden sector, which effectively regulates the spread of the effective Yukawa couplings. 
The emerging left-right symmetric framework recovers additional appealing features typical of these models, allowing for instance to identify the hypercharges of the involved fermions with their $B-L$ charges and offering a straightforward solution to the strong CP problem.
The scheme gives rise to a distinguishing phenomenology that potentially can be tested at the LHC and future colliders through the same interactions that result in the radiative generation of Yukawa couplings, as well as by exploiting the properties of the additional $SU(2)_R$ Higgs doublet.
\newpage

\section{Introduction} 
\label{sec:Introduction}

The gauge structure of the Standard Model (SM) requires the presence of Yukawa couplings between the Higgs doublet and fermions to ensure the gauge invariance and renormalizability of the fermion mass generation mechanism. The recent discovery at the LHC of a Higgs boson \cite{Aad:2012tfa} with properties well in agreement with the SM expectations has demonstrated that the Higgs mechanism \cite{Englert:1964et} for the electroweak and chiral symmetry breaking is realised in Nature and, coincidently, strengthened our belief in the fundamental nature of SM Yukawa couplings. However, unlike the case of gauge interactions, these quantities are not connected with any symmetry principle and their observed values at low energy seem to obey no scheme. More in detail, the unusual span of the Yukawa couplings, reflected in the large differences measured amongst the masses of fundamental fermions, remains one of the few unknown aspects of the SM and poses the so-called flavour hierarchy puzzle. In absence of a proven mechanism to legitimate the fundamental nature of Yukawa couplings, it is therefore mandatory to investigate the flavour hierarchy problem within a broad set of frameworks, including also constructions which propose an effective origin of these parameters.
The most popular approach in this direction is probably that of Froggat and Nielsen \cite{Froggatt:1978nt}. The scheme is based on a $U(1)_F$ global symmetry associated to the existence of flavour symmetric non-renormalizable local operators with very high dimensionality, that involve large powers of scalar flavon fields. Although the mechanism can elegantly explain the hierarchy of Yukawa couplings in terms of higher powers of the flavons vacuum expectation values (VEVs), it is still not clear what kind of underlying physics can be responsible for the generation of such operators and whether the scheme is actually testable.

In alternative to that, a more recent solution \cite{Gabrielli:2013jka} introduced a new mechanism which produces exponentially spread effective Yukawa couplings at low energy owing to the dynamics of a hidden sector. Such mechanism is based on the hypothesis that Yukawa couplings, which vanish at the tree-level by requirement of a new $Z_2$  symmetry, are radiatively generated after the spontaneous symmetry breaking (SSB) of the latter. Given that the theory is renormalizable, the loop diagrams behind the Yukawa coupling generation are necessarily finite at any order in perturbation theory and the scheme is thus predictive. In the model proposed in \cite{Gabrielli:2013jka}, the main source of chiral and flavour symmetry breaking is identified with the Dirac masses of dark fermions: replicas of SM fermions which are singlet under the gauge group of the latter. The chiral and flavour symmetry breaking is then communicated at the 1-loop level by a set of scalar messenger fields, which carry here the same quantum numbers of squarks and sleptons from supersymmetric extensions of the SM. The dark fermions, together with the scalar messangers, form the hidden sector of the theory, the purpose of which is to induce and transmit the chiral symmetry breaking to the SM. In this regard, notice that the choice of a mechanism to forbid the presence of Yukawa couplings at the tree level in such a framework is indeed not unique. Furthermore, other setups for the hidden sector can successfully accomplish and transfer the necessary symmetry breaking, as shown for instance in \cite{Davidson:1987mh, Brahmachari:2003wv}, within the context of neutrino physics in  \cite{Ma:2014rua} and in \cite{preparation} within a new supersymmetric scenario.
  
Fascinated by the idea of an effective origin of Yukawa couplings, in the present paper we adopt the setup of \cite{Gabrielli:2013jka} as a prototypal  hidden sector and show that Yukawa couplings can be elegantly forbidden at the tree level by gauge symmetry solely, by extending the SM gauge group to the simplest realization of the left-right (LR) symmetric group $SU(2)_L\times SU(2)_R \times U(1)_Y$. More in detail, we consider in our scheme a Higgs sector composed by a new $SU(2)_R$ Higgs doublet in addition to the usual SM-like $SU(2)_L$ SM Higgs field. The vanishing of tree level Yukawa couplings is then guaranteed by the absence of Higgs fields in the bi-doublet representation of the LR group, typically included in this kind of model \cite{Mohapatra:1974gc,Mohapatra:1974hk,Senjanovic:1975rk,Mohapatra:1977mj,Senjanovic:1978ev,Duka:1999uc,Tello:2010am,Maiezza:2010ic} precisely to have fundamental gauge-invariant Yukawa couplings.  
In this simple setup, the generation of Yukawa couplings necessarily involves higher dimensional operators; the lowest-order gauge-invariant suitable ones that we identify are 
\bea
O_{Y_f}&=&
\frac{1}{\Lambda^f_{\rm eff}}(\bar{\psi}^f_L H_L)(H^{\dag}_R \psi^f_R) + h.c. \, ,
\label{OY}
\eea
where $(\psi^f_L) \, H_L$ and $(\psi^f_R)\, H_R$ are the (fermions) Higgs doublets of the $SU(2)_L$ and $SU(2)_R$ sectors respectively, and $\Lambda^{f}_{\rm eff}$ is the corresponding effective scale computed from the fundamental theory.
After the SSB of $SU(2)_R$, operated by the $H_R$ doublet, the Yukawa operators arise from Eq.~(\ref{OY}) through the substitution $H_R \to v_R/\sqrt{2}$, and the generated SM Yukawa coupling $Y_f$ of a generic fermion $f$ is given by 
\bea
Y_f \sim \left(\frac{M_f v_R }{\bar{m}^2}\right)\times ~ {\rm loop}{\rm -}{\rm function} \, .
\label{Yuk}
\eea 
Here the parameters $\bar{m}$ and $M_f$ are related to the masses of the particles from the hidden sector that run in the loop and completely determine the emerging Yukawa couplings. 

The focus of the present paper is on the simple LR symmetric scheme proposed, identified here as an ideal embedding for hidden sectors dedicated to the radiative generation of Yukawa couplings. The purpose of our investigation is to provide analytic expressions for the properties of the new gauge bosons and assess the phenomenological viability of the framework, independently of the details of the hidden sector. In our analysis we find that all the SM observables of the EW sector are affected by corrections of order ${\cal O}((v_L/v_R)^2)$, being $v_L$ and $v_R$ the VEVs of the corresponding Higgs fields, which are completely negligible in light of the latest collider bounds on additional massive gauge bosons. Interestingly, such corrections are also in line with the hidden sector proposed in \cite{Gabrielli:2013jka}, for which the VEV of the right-handed (RH) Higgs doublet is naturally well above the 10 TeV scale. Furthermore, in spite of the atypical configuration adopted for the Higgs sector, we find that our scenario still exhibits the typical welcomed features of LR symmetric models. For instance, as we will show below, our scheme allows for the identification of the hypercharge gauge group with the $B-L$ one. On top of that, the strong CP problem can also be solved in remarkably straightforward way \cite{Mohapatra:1978fy} owing to the natural cancellation of the relevant CP violating phase which enters the determinant of fermion mass matrices \cite{Diaz-Cruz:2016pmm}. The introducing of an ad-hoc $U(1)$ Peccei-Quinn symmetry \cite{Peccei:1977hh} and the related axion field \cite{Weinberg:1977ma, Wilczek:1977pj} is therefore unnecessary.

The paper is organized as follows: in section II we present the results 
for the SM extended gauge sector and detail the symmetry breaking mechanism. The adopted Higgs sector is discussed in isolation in section III, whereas in section IV we analyze the radiative generation of Yukawa couplings in the setup of \cite{Gabrielli:2013jka}. More details on this point and sketches of other suitable hidden sectors are presented in the Appendix. In section V we discuss some phenomenological aspects of the model and finally report our conclusions in section VI.

\section{The gauge sector} 
\label{sec:The gauge sector}

We consider a framework based on the $ SU(2)_L\times SU(2)_R \times U(1)_Y $ symmetry group. The SM particle content is then extended to the RH
Higgs boson doublet $H_R$ and the RH fields are organised in doublets in the same fashion as the left-handed (LH) ones. 
In particular, the hypercharges of RH doublets are set to the corresponding LH values and we impose that the $SU(2)$ couplings of the two chiral sectors obey $g_L=g_R=g $ in order to respect the left-right symmetry. As we show below, this setup reproduces the EWSB sector of the SM up to corrections of order $(v_L/v_R)^2$.

The Lagrangian responsible for the gauge symmetry breaking 
contains the following terms
\bea
\LG & \supset (D_\mu H_R)^\dg (D^\mu H_R) = (H_{uR}^*, H_{dR}^*)
\modu{\left[g\sum_i\frac{\tau^i}{2}W_{R\mu}^i + g' Y B_\mu \right]}^2 
\begin{pmatrix}
H_{uR}\\ H_{dR}
\end{pmatrix},
\eea
where $ \tau^i $, $i=\{1,2,3\}$, are the $ SU(2)_R $ generators and $ Y $ the hypercharge operator.
After both the $ SU(2)_{L/R }$ symmetry breaking, arising from the VEVs of the Higgs fields 
\begin{equation}
	\left\langle H_{L/R}\right\rangle \to \begin{pmatrix}
		0 \\  v_{L/R}/\sqrt 2
	\end{pmatrix}\, ,
	\label{eq:LRbr}
\end{equation} 
the  Lagrangian for the mass term of the gauge sector reduces to
\bea
\LG &\supset& \frac{g^2}{4}\left(v^2_R\, W^{\pm \dg}_R W^{\pm}_R + v^2_L \, W_L^{\pm \dg} 
W_L^{\pm}\right) + \frac{v_R^2}{8} \left[g \,  W_R^3 - g' B \right]^2 + \frac{v_L^2}{8} \left[g \,  W_L^3 - g' B \right]^2,
\eea
where saturated Lorentz indices are left understood. The squared masses of the charged gauge bosons $W_L^{\pm}$ and $W_R^{\pm}$, associated to the $ SU(2)_{L}$ and $ SU(2)_{R }$ gauge group respectively, are given by
\bea
\operatorname{M}_{W_{L,R}}^2&=&\frac{g^2 v^2_{L,R}}{4}\, .
\label{MW}
\eea
In the neutral gauge boson sector, the above Lagrangian gives rise to the following squared mass matrix
\bea
\operatorname{M}^2 &=& 
\frac 1 8 \begin{pmatrix}
(v_L^2 + v_R^2) g'^2 & -v_L^2g g' & -v_R^2g g' \\
-v_L^2g g' & v_L^2 g^2 & 0 \\
-v_R^2g g' &0 & -v_R^2g^2 
\end{pmatrix}
\label{MassN}
\eea
on the basis $ (B, W^3_L, W^3_R)^T$.

In line with the latest results of collider experiments, we require throughout the paper the VEV hierarchy $v_R\gg v_L $. This allows us to analyse the mass eigenstates and eigenvectors of the neutral gauge boson mass matrix in Eq.~(\ref{MassN}) by retaining only the leading corrections in the $v_L/v_R$ expansion. 

By defining $ (A, Z_L, Z_R)$ the basis where the matrix $\operatorname{M}^2 $ is diagonal, we find, up to terms of order $ \mathcal{O}((v_L/v_R)^{2})$, that
\bea
M_A & = & 0, \\
\label{MZL}
M_{Z_L}^2 & =& \frac{(g^4 + 2 g'^2 g^2)v_L^2}{4(g'^2+g^2)} ,\\
\label{MZR}
M_{Z_R}^2 & =& \frac{(v_L^2 + v_R^2)g'^4  + 2 g'^2 g^2 v_R^2 + g^4 v_R^2}{4(g'^2+g^2)}\, ,
\eea
with the corresponding eigenstates
\bea
A_\mu & =& \frac{g B_\mu + g'(W^3_{L\mu}+W^3_{R\mu})}{\sqrt{2 g'^2 + g^2}} ,\\
\label{ZL}
Z_{L\mu} &=&\frac{ g'^2(W^3_{L\mu} - W^3_{R\mu})- g g'  B_\mu + g^2 W^3_{L\mu}}{\sqrt{2 g'^4 +3 g^2g'^2+g^4}} ,\\
\label{ZR}
Z_{R\mu} &=&  \frac{gW^3_{R\mu} - g' B_\mu }{\sqrt{ g'^2 + g^2}} \, .
\eea
As we can see, the massless state $A_\mu$ corresponds to the standard photon field,  whereas the heavy neutral gauge bosons $Z_L$ and  $Z_R$ are to be identified with the usual $Z^{0}$ of the SM and its new counterpart associated to the $ SU(2)_R $ group.

In the considered limit, the inverse relations are
\bea
B_\mu &=& \frac{gA_\mu }{\sqrt{2g'^2+g^2}}-\frac{g' Z_{R\mu}}{\sqrt{g'^2+g^2}}-\frac{g' g Z_{L\mu}}{\sqrt{2 g'^4+3 g'^2 g^2+g^4}},\\
\noindent
W^3_{L\mu} & =& \frac{ g' A_\mu}{\sqrt{2 g'^2+g^2}}+\frac{Z_{L\mu}\left(g'^2+g^2\right)}
{\sqrt{2 g'^4+3g'^2g^2+g^4}},\\
\noindent
W^3_{R\mu}&=&\frac{g'A_\mu}{\sqrt{2 g'^2+g^2}}+\frac{gZ_{R\mu}}{\sqrt{g'^2+g^2}}-\frac{g'^2 Z_{L\mu}}{\sqrt{g^4+3 g^2g'^2+2g'^4}},
\label{basis}
\eea
which we can use to rewrite the Lagrangian in terms of the fields corresponding to the mass eigenstates. Considering the covariant derivative that acts on a generic Dirac fermion $ \Psi=\Psi_L+\Psi_R$, we obtain
\bea
\LG &\supset& 
 \bar{\Psi}_L \cancel{D} \Psi_L + \bar{\Psi}_R \cancel{D} \Psi_R 
\supset   i\bar{\Psi}_L\left(g\cancel{W^3_L }I^3_L+ g' \cancel{B}Y \right)  \Psi_L + (L\leftrightarrow R) = \\
& =&
  i\bar{\Psi}_L\!\!\left[
\frac{g' g }{\sqrt{2g'^2+g^2}}(I^3_L+Y)\cancel{A}
+
\frac{g\left[g'^2( I^3_L-Y) +g^2I^3_L\right]}{\sqrt{\left(g'^2+g^2\right) \left(2g'^2+g^2\right)}}\cancel{Z_L}
-
\frac{g'^2 }{\sqrt{g'^2+g^2}}Y\cancel{Z_R}
\right]\!\!\Psi_L \\
& +&  i\bar{\Psi}_R\!\!\left[
\frac{g' g }{\sqrt{2g'^2+g^2}}(I^3_R+Y)\cancel{A}
-
\frac{g g'^2 (I^3_R+Y)}{\sqrt{\left(g'^2+g^2\right) \left(2 g'^2+g^2\right)}}\cancel{Z_L}
+
\frac{g^2 I^3_R -g'^2 Y }{\sqrt{g'^2+g^2}}\cancel{Z_R}
\right]\!\!\Psi_R. 
\eea
By analysing the coupling of photons to the chiral $L/R$ multiplets we can identify 
\bea
Q_{L/R} & =& I_{3}^{L/R}+Y,\\
e & =& \frac{g' g }{\sqrt{2g'^2+g^2}} = g\sin\tw= g'\sqrt{\cos2\tw},
\eea
and remark that there is no way to test the above rightmost equality because $ g'$ is not measured, although in principle this could discriminate between the current framework and that of the SM, where $e=g'\cos\tw^{SM} $. Notice also that the requirement $ Q_L=Q_R$ forces the 
LH and RH components of a given field to have the same hypercharge, namely $Y(\Psi_L)=Y(\Psi_R)$, and the general relation 
between the charge $Q$ and hypercharge $Y$ operators becomes
\bea
Q & =& I_{3}^{L}+I_{3}^{R}+Y\, ,
\eea
with $I_{3}^{L}(\Psi_{R})=I_{3}^{R}(\Psi_{L})=0$. We can then rewrite the interacting Lagrangian of neutral currents as
\bea
\LG_{\rm NC} & =& 
  ieQ_\Psi\left[\bar{\Psi}\cancel{A}\Psi\right]
  +  i\frac{g}{\cos\tw} \left[\bar{\Psi}\cancel{Z_L}
 \left( I^3_\Psi \PL - \sin^2\tw Q
\right) \Psi \right] \\
\nonumber
&  + &  i g\frac{\sqrt{\cos2\tw}}{\cos\tw} \left[\bar{\Psi} \cancel{Z_R} \left( I^3_\Psi \PR -  \frac{\sin^2\tw}{\cos2\tw}(Q_\Psi -   I^3_\Psi) \right)\Psi\right]\,  +{\cal O}((v_L/v_R)^2)\, ,
\eea
where $\PL=(1-\gamma_5)/2$, $\PR=(1+\gamma_5)/2$ and the third component of 
the Isospin operator $I^3_\Psi$ is now acting on the $SU(2)$ multiplet $\Psi$.

Notice that by writing explicitly the hypercharge assignments of the fermions in the present framework, reported in Table~\ref{tab0}, we can identify these quantities with the $B-L$ charges of the involved fields and write
\begin{equation}
	Y \equiv \frac{B-L}{2},
\end{equation}
where we used the fact that the baryon number of a quark is 1/3 whereas the lepton number of a lepton is 1. This is a generic feature of LR frameworks which nicely explains the lack of anomalies in the accidental $U(1)_{B-L}$ symmetry of the SM.  

\begin{table}[h]
	\centering
	\begin{tabular}{cccc}
	\toprule
	Field & \(I^3_X\) & \(Y\) & \(Q=I_3+Y\)\tabularnewline
	\midrule
	\(u_L\) & 1/2 & 1/6 & 2/3\tabularnewline
	\(d_L\) & -1/2 & 1/6 & -1/3\tabularnewline
	\(u_R\) & 1/2 & 1/6 & 2/3\tabularnewline
	\(d_R\) & -1/2 & 1/6 & -1/3\tabularnewline
	\(e_L\) & -1/2 & -1/2 & -1\tabularnewline
	\(\nu_L\) & 1/2 & -1/2 & 0\tabularnewline
	\(e_R\) & -1/2 & -1/2 & -1\tabularnewline
	\(\nu_R\) & 1/2 & -1/2 & 0\tabularnewline
	\bottomrule
	\end{tabular}
	\caption{Hypercharge and $I^3_X$, $X=L,R$, assignments of the SM fermions in the present framework.}
	\label{tab0}
\end{table}

\section{ Higgs Sector}
\label{sub:Higgs}

The Lagrangian for the Higgs sector of the theory is
\bea
	\LG \supset \LG_H &=& (D_\mu H_L) (D^\mu H_L)^\dg 
	+ 
	(D_\mu H_R) (D^\mu H_R)^\dg 
	- 
	\lambda_L (H_L H_L^\dg)^2 
	-
	\lambda_R (H_R H_R^\dg)^2
	\\ \nn
	& - &
	\lambda_{LR}(H_R H_R^\dg)(H_L H_L^\dg)
	+
	\mu_L^2 (H_L H_L^\dg)
	+
	\mu_R^2 (H_R H_R^\dg)\, ,
	\label{eq:lagH}
\eea
where $H_{L,R}$ are the corresponding $SU(2)_{L,R}$ doublets. The coefficients of the quartic and portal terms have been chosen in a way that, after both $SU(2)_{L,R}$ 
symmetry breaking, the corresponding interaction terms are normalised by a factor of 1/4:
\bea
	\LG_H &\to &\frac{1}{2} (\pd_\mu h_L) (\pd^\mu h_L)^\dg 
	+ 
	\frac{1}{2} (\pd_\mu h_R) (\pd^\mu h_R)^\dg 
	- 
	\frac{1}{4} 
	\lambda_L (h_L + v_L )^4 
	\\ \nn
	& - &
	\frac{1}{4}
	\lambda_R (h_R + v_R)^4
	-
	\frac{1}{4}
	\lambda_{LR}(h_R +v_R)^2(h_L + v_L)^2
	+
	\frac{1}{2}
	\mu_L^2 (h_L+ v_L)^2
	+
	\frac{1}{2}
	\mu_R^2 (h_R+ v_R)^2\, .
\eea
In the following we will consider a minimal model in which the $SU(2)_L$ symmetry breaking arises from the spontaneous $SU(2)_R$ breaking via a (negative) portal coupling $\lambda_{LR}$. We then retain a (negative) tree-level mass term for the RH Higgs doublet, $\mu^2_R >0 $ but set $\mu^2_L = 0 $. The emerging scenario therefore capture the gist of scale invariant extension of the model, in which $\mu^2_R >0 $ is generated for instance through the Coleman-Weinberg mechanism. 

\subsection{Positivity of the scalar potential}
\label{ssub:positivity}
As far as the stability of the potential in the Higgs sector is concerned, the negative mass term of $H_R$ has a negligible effect compared to the other terms in the potential once large values of the fields are assumed. Therefore, the
positivity of the potential will depend only on the quartic and
portal couplings, exactly as in a simplified version of the 2HDM where
only $\lambda_{i\leq3}\neq 0$  \cite{Branco:2011iw}. Hence, by a direct comparison with Eq.s (98) and (176) of Ref.\cite{Branco:2011iw}, we have that the potential is bounded from below if
\bea
 \lambda_{L,R} &\geq& 0,
\nonumber \\
 \lambda_{LR}&\geq&-2\sqrt{\lambda_L\lambda_R}\, .
\eea

\subsection{Minimisation conditions and mass
eigenstates}
\label{ssub:eigenstates}

The minimum of the potential is obtained as usual by requiring that the gradient of potential vanish for $H_{L/R}=v_{L/R}/\sqrt 2$ or, analogously, by setting the 
gradient of the manifestly broken potential to vanish for $h_X=0$. Explicitly
\bea
	\nabla V(h_L, h_R) \Big\vert_{h_X=0}=0\implies v_L=v_R=0\quad\lor\quad
	\begin{cases}
	\lambda_L v_L^2 +\dfrac{1}{2}\lambda_{LR}v_R^2=0
	\\
	\lambda_R v_R^2 +\dfrac{1}{2}\lambda_{LR}v_L^2 - \mu_R^2=0
	\end{cases}\, ,
\eea
where the $v_L=v_R=0$ solution corresponds to the local maximum of the potential. Notice that, the usual breaking relation for the right sector,  $v_R = \mu^2_R/\lambda_R$, is correctly recovered for \(\lambda_{LR}=0\). Focusing on nonzero values of the VEVs we obtain
\bea
\label{eq:mincon}
	\begin{cases}
	v_L^2 =- \dfrac{\lambda_{LR}}{2\lambda_L }v_R^2=0 \quad \implies\quad v_L\in \mathbb{R} \Longleftrightarrow \lambda_{LR}<0,
	\\
	 v_R^2 \left(\lambda_R - \dfrac{\lambda_{LR}^2}{4\lambda_L }\right) =  \mu_R^2\quad \implies\quad v_R\in \mathbb{R} \Longleftrightarrow 2\sqrt{\lambda_L\lambda_R}>\lambda_{LR}> -2\sqrt{\lambda_L\lambda_R},
	\end{cases}
\eea
with the second condition being actually stronger than the one imposed
by the positivity of the potential. Therefore, the conditions for the
consistency of the model are summarised in
\begin{equation}
	\begin{cases}
		-2\sqrt{\lambda_L\lambda_R} < \lambda_{LR}< 0,\\
		\nonumber
		\lambda_{L,R} \geq 0\, .
	\end{cases}
\end{equation}
The first condition implies, through the top relation in Eq.~\eqref{eq:mincon}, the following hierarchy in the VEVs
\bea
\frac{v_L^2}{v_R^2} &<& \sqrt{\frac{\lambda_R}{\lambda_L}}\, .
\label{eq:cond}
\eea

To obtain the mass matrix for the two Higgs fields we compute the second
derivative of the potential after the symmetry breaking, evaluated for 
vanishing values of the fields, obtaining
\bea
\ON{M}^2=
2v_R^2
\begin{pmatrix}
\lambda_L\epsilon^2 & -\lambda_L\epsilon^3\\
-\lambda_L\epsilon^3&\lambda_R
\end{pmatrix}\, ,
\eea
where we defined $\epsilon:=v_L/v_R$ and used
\bea
\lambda_{LR}&=&-2\lambda_L \left(\frac{v_L}{v_R}\right)^2;\qquad\mu^2_R = \frac{\lambda_R v_R^4 - \lambda_L v_L^4}{v_R^2}\, ,
\eea
which can be derived from Eq.~\eqref{eq:mincon}. Notice that, by
imposing the positivity of the determinant of $\ON{M}^2$, one recovers
the condition in Eq.~\eqref{eq:cond}. Notice also that the off diagonal terms
in the above matrix are of higher order than the diagonal ones, hence at
the lowest order in the limit $v_L/v_R \ll1$ the eigenvalues coincide with the
diagonal elements and the above condition on the squared ratio of the VEVs
translates into
\bea
m_1^2&<&\sqrt{\frac{\lambda_L}{\lambda_R}}m_2^2 \, .
\eea
By considering higher order terms in the $\epsilon$ expansion, we have
\begin{align}
m_1^2 & = 2  \lambda_L  \epsilon^2  v_R^2 \left(1 - \frac{\lambda_L}{\lambda_R}\epsilon^4\right) + \mathcal{O}(\epsilon^8),
\\ 
m_2^2 &= 2 \lambda_R v_R^2 \left(1  + \frac{\lambda_L^2}{\lambda_R^2}\epsilon^6\right)+ \mathcal{O}(\epsilon^8),
\end{align}
which can be obtained by diagonalising the original mass matrix with the following mixing matrix 
\bea
\ON{U}&=&
\begin{pmatrix}
1-\dfrac{\lambda^2_L\epsilon^6\left(2\lambda_L\epsilon^2+\lambda_R\right)}{2\lambda_R^3}
&
-\dfrac{\lambda_L\epsilon^3\left(-\lambda_L^2\epsilon^4 + 2\lambda_L\lambda_R\epsilon^2 + 2\lambda_R^2\right)}{2\lambda_R^3}
\\
\dfrac{\lambda_L\epsilon^3\left(-\lambda_L^2\epsilon^4 + 2\lambda_L\lambda_R\epsilon^2 + 2\lambda_R^2\right)}{2\lambda_R^3}
&
1-\dfrac{\lambda^2_L\epsilon^6\left(2\lambda_L\epsilon^2+\lambda_R\right)}{2\lambda_R^3}
\end{pmatrix}\, ,
\eea
satisfying
\beq
\ON{U}^T \ON{M^2}\ON{U}=\text{diag}(m_1^2, m_2^2)+\mathcal{O}(\epsilon^8);
	\qquad
	\lim_{\epsilon\to0}\ON{U}=\mathbb{I};
	\qquad 
	\ON{U}^T\ON{U}=\ON{U}\ON{U}^T=\mathbb{I}+\mathcal{O}(\epsilon^9)\, .
\eeq
The expression for the original gauge eigenstates in term of the mass
eigenstates can therefore be obtained as
\bea
\begin{pmatrix}
h_L \\ h_R
\end{pmatrix}
=
\ON{U}
\begin{pmatrix}
h_1 \\ h_2
\end{pmatrix},
\eea
and, at the lowest order, we have $h_L\approx h_1$, $h_R \approx h_2$.

\section{Radiative Yukawa couplings} 
\label{sec:Radiative Yukawa couplings}

As made clear in the Introduction, the radiative generation of Yukawa couplings requires a hidden sector of the theory to source the chiral symmetry breaking. As a first example, we adopt here the setup originally proposed in \cite{Gabrielli:2013jka}, consisting in a set of dark massive Dirac fermions singlet under the  $SU(2)_L \times SU(2)_R\times U(1)_Y$ group, but charged under an unbroken $U(1)_F$ gauge interaction. We then make use of
a non-perturbative mechanism, based on the solution of the gap-equation via
the Nambu-Jona-Lasion mechanism \cite{Nambu:1961tp},
to generate the exponential spread in the dark fermion mass splitting \cite{Gabrielli:2007cp,Gabrielli:2014oya}. This requires 
the existence of a a higher derivative term in the pure $U(1)_F$  gauge sector, which can be associated to the presence of a massive Lee-Wick ghost in the spectrum \cite{Lee:1971ix,Grinstein:2007mp}. The chiral symmetry breaking in the hidden sector, encoded in the dark-fermion masses, is finally transmitted to the SM through a set of scalar messenger fields which interact with both the dark and SM particles. Gauge invariance of such couplings forces the messenger fields to carry the same quantum numbers as squarks and sleptons of the SM supersymmetric extensions, leading to interesting phenomenological implications at the LHC and future $e^+e^-$ linear colliders investigated for instance in \cite{Gabrielli:2014oya}.

To embed the mechanism of \cite{Gabrielli:2013jka} in our framework, we separate the full Lagrangian into four sectors 
\bea
{\cal L}&= {\cal L}^{ Y=0}_{ESM} + {\cal L}_{DS}+ {\cal L}_{MS} + {\cal L}_{VH},
\label{totlagr}
\eea
being now ${\cal L}^{Y=0}_{ESM}$ the LR symmetric SM Lagrangian with vanishing tree-level Yukawa couplings,  ${\cal L}_{DS}$ the hidden sector Lagrangian containing the dark-fermions interactions and ${\cal L}_{MS}$ describing the scalar messenger sector with the relative interactions involving SM fermion fields, gauge fields and dark fields. Finally, ${\cal L}_{VH}$ contains the potential for the Higgs fields, which operate the $SU(2)_L\times SU(2)_R$ symmetry breaking.
Further details regarding the ${\cal L}_{DS}$ Lagrangian
and the $U(1)_F$ gauge sector are presented in \cite{Gabrielli:2013jka} and in the Appendix, which contains examples of non-perturbative dynamics that result in the desired dark-fermion spectrum.

\subsection{The messenger sector}
The  ${\cal L}_{MS}$ Lagrangian in Eq.~(\ref{totlagr}) includes here a set of new messenger doublets transforming under the $SU(2)_R$ and can be split into two terms
\bea
{\cal L}_{MS}&=&{\cal L}^{\rm 0}_{MS}+{\cal L}^{\rm I}_{MS}\, ,
\label{MSMS}
\eea
where ${\cal L}^{\rm 0}_{MS}$ contains the kinetic terms of the 
messenger fields, comprehensive of the interactions with the SM gauge fields, while ${\cal L}^{\rm I}_{MS}$ provides the messenger interactions with SM fermions, dark fermions, and the Higgs boson, which are responsible for the radiative generation of Yukawa couplings. 

The quarks quantum numbers set the minimal matter content needed for the colored messenger scalar sector, which in the case of the $SU(2)_L\times SU(2)_R\times U(1)_Y$ gauge group is given by

\begin{itemize}
\item $2N$ complex scalar $SU(2)_L$ doublets: $\hat{S}_L^{\U_i}$ and $\hat{S}_L^{\D_i}$,
\item $2N$ complex scalar  $SU(2)_R$ doublets: $\hat{S}_R^{\U_i}$ and $\hat{S}_R^{\D_i}$,
\end{itemize} 

where
$\hat{S}_{A}^{\U_i,\D_i}=\left(\begin{array}{c}S^{\U_i,\D_i}_{A,1}\\S^{\U_i,\D_i}_{A,2}
\end{array}\right)$,  with $A=\{L,R\}$, 
$N=3$ and $i=1,2,3$ stands for a family index. The $L,R$ labels identify here the messengers which couple to the $L,R$ chirality structure of the associated SM fermions (in the same fashion as for the squark fields in supersymmetric theories).
The $\hat{S}_{L,R}^{\U_i,\D_i}$, fields carry the SM quark quantum numbers corresponding to the same chirality $L,R$ of SM fermions and interact with the electroweak gauge bosons and gluons via their covariant derivatives.  
As minimal flavour violation hypothesis requires this Lagrangian be invariant under $SU(N_F)$, where $N_F$ is the number of
flavours, we can reduce the messenger mass sector to 4 different  
universal mass terms in both the $\hat{S}_{L,R}^{\U_i}$ and $\hat{S}_{L,R}^{\D_i}$ sectors. Notice that the alternative minimal hypothesis presenting a common scalar mass for the $L$ and $R$ scalar sector is also phenomenologically acceptable.

We do not report here the expression for the interaction Lagrangian of the messenger fields with the SM gauge bosons, which follow from the universal properties of gauge interactions. Notice that each messenger field is  also charged under $U(1)_F$ and carries the same $U(1)_F$ charge as the associated dark fermion, hence $U(1)_F$ charges identify the flavour state.
A summary of the relevant quantum numbers of the new strongly-interacting (according to QCD) fermion and scalar fields can be found in Table~\ref{tab1}.

\begin{table} \begin{center}    
\begin{tabular}{ccccccc}
\toprule 
Fields 
& Spin
& $SU(2)_L$ 
& $SU(2)_R$ 
& $U(1)_Y$
& $SU(3)_c$
& $U(1)_F$
\tabularnewline \midrule 
$\hat{S}_L^{\D_i}$
& 0
& 1/2
& 0
& 1/3
& 3
& -$\qdi$
\tabularnewline 
$\hat{S}_L^{\U_i}$
& 0
& 1/2
& 0
& 1/3
& 3
& -$\qui$
\tabularnewline 
$\hat{S}_R^{\D_i}$
& 0
& 0
& 1/2
& 1/3
& 3
& -$\qdi$
\tabularnewline 
$\hat{S}_R^{\U_i}$
& 0
& 0
& 1/2
& 1/3
& 3
& -$\qui$
\tabularnewline 
$Q^{\D_i}$
& 1/2
& 0
& 0
& 0
& 0
& $\qdi$
\tabularnewline 
$Q^{\U_i}$
& 1/2
& 0
& 0
& 0
& 0
& $\qui$
\tabularnewline \bottomrule \end{tabular} 
\caption{
Spin and gauge quantum numbers of the strongly interacting messenger fields and 
corresponding dark fermions.   
$U(1)_F$ is the gauge symmetry associated to the dark photon of the hidden sector.
}
\label{tab1}
\end{center} \end{table}

In order to generate radiative Yukawa couplings which are gauge invariant under all interactions, we take the following expression for the ${\cal L}^I_{MS}$ Lagrangian
\bea
{\cal L}^I_{MS} &=&
\hat{g}_L   \left( \sum_{i=1}^{N}\left[\bar{q}^i_L Q_R^{\U_i}\right] \hat{S}^{\U_i}_{L} +
\sum_{i=1}^{N}\left[\bar{q}^i_L Q_R^{\D_i}\right] \hat{S}^{D_i}_{L}\right)
\nonumber\\
&+&
\hat{g}_R \left(\sum_{i=1}^{N}\left[\bar{q}^i_R Q_L^{\U_i}\right] \hat{S}^{\U_i}_{R} +
\sum_{i=1}^{N} \left[\bar{q}^i_R Q_L^{\D_i}\right] \hat{S}^{\D_i}_{R}\right) 
\nonumber\\
&+& \;
\lambda \sum_{i=1}^{N}\left(\tilde{H}_L^{\dag} \hat{S}^{\U_i}_L \hat{S}^{\U_i\dag}_R\tilde{H}_R
+ H^{\dag}_L \hat{S}^{\D_i}_L \hat{S}^{\D_i\dag}_RH_R \right) 
\,+\, h.c.  \,,
\label{LagMS}
\eea 
where contractions with color and $SU(2)_{L,R}$ indices are left understood. The $SU(2)_{L,R}$ doublets $q^i_{L,R}=\left(\begin{array}{c} U^i_{L,R}\\D^i_{L,R} \end{array}\right)$
represent here the SM up ($U$) and down ($D$) quark fields,  
$H_{L,R} =\left(\begin{array}{c} H_{L,R}^{\pm}\\H^{0}_{L,R}  \end{array}\right)$
are the Higgs doublet and $\tilde{H}_{L,R}$ are defined as
$\tilde{H}_{L,R}=i\sigma_2 H^{\star}_{L,R}$. 
Notice that the above interacting Lagrangian gives rise to diagonal Yukawa couplings in the weak interaction basis for both up and down quark fields. A generalization of the scenario that correctly reproduces the measured CKM mixing can be found in \cite{Gabrielli:2016cut}.
The $\hat{g}_L$ and $\hat{g}_R$ constants in Eq.~\eqref{LagMS} are flavour-universal free parameters whose values remain in the perturbative regime $\hat{g}_{L,R}< 1$. In the following we will identify $\hat{g}_L=\hat{g}_R=\g$ as required by LR symmetry.

\subsection{The origin of Yukawa couplings}
Consequently to the breaking of $SU(2)_L\times SU(2)_R$, Eq.~\eqref{eq:LRbr}, the Lagrangian of the $S_{L,R}$ fields becomes 
\bea
{\cal L}^0_{S}&=& \partial_{\mu} \hat{S}^{\dag} \partial^{\mu}\hat{S} - \hat{S}^{\dag} M^2_S \hat{S},
\eea
where $\hat{S}\equiv (\hat{S}_L,\hat{S}_R)$ and the square mass term is given by
\begin{equation}
M^2_S = \left (
\begin{array}{cc}
m^2_L & \Delta \\
\Delta & m^2_R 
\end{array}
\right),
\label{M2}
\end{equation}
with $\Delta=\frac{1}{2}\lambda v_R v_L$ parametrising the left-right scalar mixing. We have omitted here the $U,D$ indices of the messenger fields since they are not relevant for the purposes of the following discussion. The full $SU(6)$ flavour universality requires the terms appearing in Eq.~(\ref{M2}) 
be constant for each $\hat{S}^{\U_i,\D_i}_{L,R}$ flavour component.
Then, for each flavour, the $M^2_S$ matrix in Eq.~(\ref{M2}) can be diagonalized by the unitary matrix
\begin{equation}
U = \left (
\begin{array}{cc}
\cos{\theta} & \sin{\theta} \\
-\sin{\theta} & \cos{\theta} 
\end{array}
\right),
\label{U2}
\end{equation}
 with  
$\tan{2\theta}=\frac{2\Delta}{m_L^2-m_R^2}$. 
The eigenvalues of the diagonal $M^{2 \,\rm diag}_S=U M^2_S \,U^{\dag}$ matrix are
\bea
m^2_{\pm}=\frac{1}{2}\left(m^2_L+m_R^2 \pm
\left[(m^2_L-m^2_R)^2+4\,\Delta^2\right]^{1/2}
\right)\, .
\eea
Since we restrict our analysis to the symmetric LR scenario, we take here $m^2_L=m^2_R=\bar{m}^2$ and the $U$ matrix elements consequently
simplify to $U(i,i)=1/\sqrt{2}, \;U(1,2)=\!-U(2,1)=1/\sqrt{2}$, yielding square-mass eigenvalues 
\bea
m^2_{\pm}=\bar{m}^2(1\pm \xi)\, ,
\eea
where the mixing parameter $\xi$ is given by 
\bea
\xi=\frac{\lambda v_R v_L}{2\bar{m}^2}\, .
\label{mixing}
\eea
Notice that the notation used here for the mixing parameter as a function of the VEVs differs from the corresponding one in \cite{Gabrielli:2013jka} by a factor 1/2, due to a different parametrizations of the VEVs.

\begin{figure}[h]
  \centering
    \includegraphics[width=.7\textwidth]{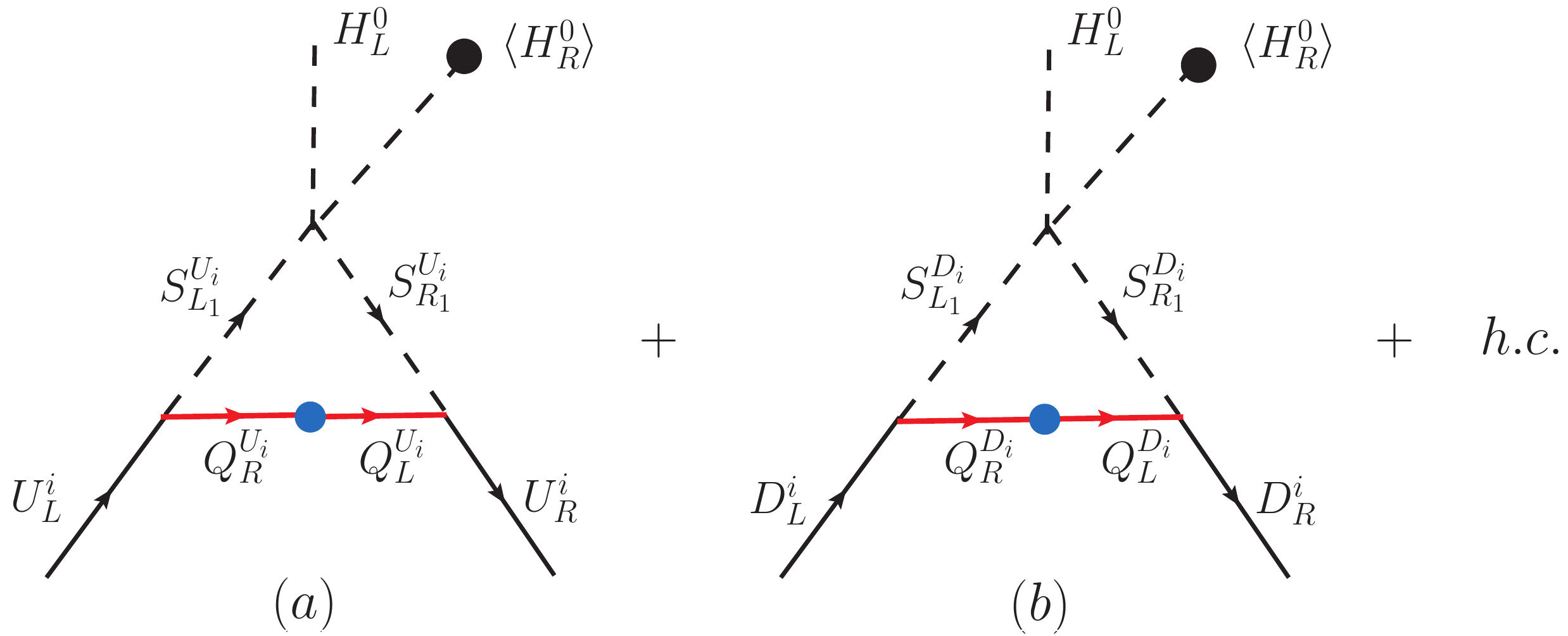}
  \caption{The diagrams responsible for the radiative generation of the Yukawa couplings of up-type quarks (a) and down-type quarks. The dot on the external Higgs line signifies that the field is set at its vacuum expectation value.}
  \label{fig1}
\end{figure}

We consider now the SM Yukawa coupling $Y_f$ of a generic fermion $f$ to the left-handed doublet $H_L$. Since the masses of the messenger fields are 
quite large in this scenario (well above the TeV scale for the quark sector), 
we can apply the results of effective field theory. In particular, due to the Feynman diagrams in Fig.\ref{fig1}, the following 
dimension 5 operator is generated at 1-loop
\bea
{\cal L}_{eff} &=& \frac{1}{\Lambda^f_{\rm eff}}(\bar{\psi}^f_L H_L)(H^{\dag}_R \psi^f_R) + h.c. \, ,
\label{OYukawa}
\eea
where $(\psi^f_L) \, H_L$ and $(\psi^f_R)\, H_R$ are the generic (fermions) Higgs doublets for the $SU(2)_L$ and $SU(2)_R$ sectors respectively, and $\Lambda^{f}_{\rm eff}$ is the corresponding effective scale, which depends on the flavour $f$ of the fermion field $\psi^f$.

Then, after the SSB of $SU(2)_R$ gauge group, Yukawa couplings arise by 
replacing the $H_R$ double with its VEV $v_R$.  Analogous results hold for the Yukawa couplings of $H_R$ and follows by replacing 
$H_L$ with $v_L$ in Eq.~(\ref{OYukawa}). It is then clear that the Yukawa couplings of the $H_R$ in this model are predicted to match the corresponding Yukawa couplings of the SM (LH) Higgs times a rescaling factor of $v_L/v_R$.

The exact expression for $Y_f$ as a function of the mixing $\xi$ is
\cite{Gabrielli:2016cut}
\bea
Y_f&=&\left(\frac{\g^2 }{16 \pi^2 }\right)
\left(\frac{\xi M_{Q_f}\sqrt{2}}{v_L}\right) f_1(x_f,\xi)\, ,
\label{Yukexact}
\eea
where $M_{Q_f}$ is the mass of the $Q_f$ associated dark-fermion, $x_f=M_{Q_f}^2/\bar{m}^2$, and the loop function $f_1(x,\xi)$ is given by 
\cite{Gabrielli:2016cut}
\bea
f_1(x,\xi)&=&\frac{1}{2}\left[
C_0(\frac{x}{1-\xi})\frac{1}{1-\xi}+C_0(\frac{x}{1+\xi})\frac{1}{1+\xi}\right]\,.
\label{f1}
\eea
The basic scalar loop function $C_0(x)$ is given by
\bea
C_0(x)=\frac{1-x\left(1-\log{x}\right)}{(1-x)^2}\, ,
\label{C0}
\eea 
where $\lim_{x\to 1}C_0(x)=1/2$, and, for small $x\ll 1$,  $C_0(x)\simeq 1+{\cal O}(x)$. Notice that, for a given $\bar{m}$ and dark-fermion mass $M_{Q_f}$, all the Yukawa couplings must vanish with $\xi\to 0$ since they are proportional to $v_R$.

We remark that in the setup presented above, all massive fermions are necessarily Dirac fermions, including neutrinos. It is however possible to extend the proposed framework to accomodate Majorana neutrinos, for instance by including massive Majorana fermion singlets under the $SU(2)_L \times SU(2)_R\times U(1)_Y$. Coupling these particles to the lepton and Higgs doublets would then result, after the complete symmetry breaking, in Majorana masses for LH and RH neutrinos.     

\section{Phenomenological aspects} 
\label{sec:Lower bounds on $v_R$}
We discuss here several aspects of the phenomenology of the proposed framework. 

\subsection{Lower bound on $v_R$} 
\label{sub:Lower bound on v_R}

By using the relation which connects the SM Yukawa coupling $Y_f$ of a fermion $f$ with its mass $m_f$, explicitly $Y_f=\sqrt{2} m_f/v_L$, from Eq.~(\ref{Yukexact}) we obtain the following prediction for the generic mass $M_{Q_i}$ of 
the dark fermion associated to the SM quark $q_i$, as a function of the  quark mass $m_i$ \cite{Gabrielli:2016cut}:
\bea
M_{Q_i}&=&m_i\left(\frac{\,16\pi^2}{\g^2 }\right)\frac{1}{\xi f_1(x_i,\xi)}\, .
\label{MQi}
\eea
Here $x_i=M_{Q_i}^2/\bar{m}^2$ and $\xi$ is the universal mixing parameter in the colored messenger sector. 
In order to include the lowest order QCD corrections in  Eq.~(\ref{MQi}), we replace the Yukawa couplings of heavy quarks (or analogously the corresponding quark mass $m_i$), as well as the $\g$ coupling constant, with the corresponding running quantities evaluated at the characteristic energy scale of the average messenger mass $\bar{m}$. However, since the coupling constant $\g$ is in principle a free parameter, we can always reabsorb the mentioned corrections into a redefinitions of these 
couplings maintaining in Eq.~(\ref{MQi}) $m_i$ at the pole mass.

In order to avoid {\it stable} messengers, the lightest messenger mass ${m}_{-}$ must be larger than the mass $M_{Q_t}$ of the heaviest dark fermion  associated to the top-quark~\cite{Gabrielli:2013jka}:
\bea
{m}_{-} \ge M_{Q_t}\, ,
\label{DMcondition}
\eea
where the equality in Eq.~(\ref{DMcondition}) corresponds, for the rescaled variable $x_t$, to the condition $x_t =1-\xi$.
Notice that the vacuum stability in the scalar sector requires  $\xi\le 1$, in order to avoid tachions in the spectrum that could induce 
color/charge--breaking minima through
the generation of non-vanishing  VEV in the messenger scalar sector ~\cite{Gabrielli:2013jka}.
Furthermore, owing to the $U(1)_F$ gauge invariance, Eq.~(\ref{DMcondition}) poses a sufficient condition to avoid the stability of all messenger fields. Gauge invariance and non-universality of the corresponding $U(1)_F$ charges guarantee instead that dark fermions are stable particles. \footnote{We remark that the heaviest dark fermions could still decay without breaking the $U(1)_F$  gauge invariance.
This mechanism requires the existence of new $U(1)_F$ charged 
dark-scalar fields coupled to dark-fermions and messenger fields only. Remarkably, as suggested in\cite{Gabrielli:2016cut}, these new fields, when weakly coupled, can naturally explain the correct CKM matrix pattern in this context, provided the $U(1)_F$ charges are multiple integers of some unity charge.}

Then, by using Eq.~(\ref{DMcondition}) in the top-quark case, it is possible to compute the lowest minimum of $\bar{m}$ as a function of $\xi$, which is given by 
\cite{Gabrielli:2016cut}
\bea
\bar{m}& \ge & m_t  \left(\frac{16 \pi^2}{\g^2}\right) F(\xi) \equiv \bar{m}^{\rm min} \,  ,
\label{mbarmin}
\eea
where $F(x)$ is 
\bea
F(x)&=&\frac{8x\sqrt{1-x}}{2x+(1-x)^2
\log{\left(\frac{1-x}{1+x}\right)}}\, ,
\eea
and where we replaced the function $f_1(x_t,\xi)$ by its corresponding 
limit for $x_t\to 1-\xi$.
In the case of small arguments $x\ll 1$,  the formula above simplifies to $F(x)\simeq 2/x+1/3 +{\cal O}(x)$,  while, for  $x\simeq 1$, one obtains $F(x)\simeq 4\sqrt{1-x}+{\cal O}((1-x)^{3/2})$. 

Then, by using the definition of the universal mixing parameter $\xi$ in Eq.~(\ref{mixing}), the lower bound on $\bar{m}$ in Eq.~(\ref{mbarmin}) can be translated
into the following lower bound on $v_R$ for fixed values of $\xi$
\bea
v_R \,\ge\, \frac{2 m_t^2}{\lambda v_L}\left(\frac{16 \pi^2}{\g^2}\right)^2 
\xi F(\xi)^2\, ,
\label{vRmin}
\eea
while for arbitrary values of $\bar{m}\ge \bar{m}^{\rm min}$, we always have
\bea
v_R\,<\, \frac{2\bar{m}^2}{\lambda v_L}\, ,
\eea
due to the stability bounds.
In the case of small $\xi \ll 1$, the lower bound in Eq.~(\ref{vRmin})
can be approximated as
\bea
v_R \,\gsim\, \frac{8m_t^2}{\lambda v_L}\left(\frac{16 \pi^2}{\g^2}\right)^2 
\frac{1}{\xi}\, ,
\label{vRmin0}
\eea
while for large $\xi \lsim 1$ it becomes
\bea
v_R \,\gsim\, \frac{32 m_t^2}{\lambda v_L}\left(\frac{16 \pi^2}{\g^2}\right)^2 
(1-\xi)\xi \, .
\label{vRmin1}
\eea
The lowest minima of $v_R$ are then obtained for the largest 
values of the coupling constant allowed by perturbation theory, $\lambda \lsim 1$. 

In Table~\ref{tab2} we report the numerical values of the lowest minima of 
$v_R$ and $W_R$ mass, as a function of $\xi$ and in unity of couplings
$\lambda\sim \g =1$. In our computation we used for the SM Higgs VEV the value 
of $v_L=246$ GeV and a top-quark pole mass $m_t=173.2$ GeV. Although, in 
Eq.~(\ref{vRmin1}) the top-quark mass should be replaced 
by its value at the messenger mass scale, we will still use its 
value at the mass pole and reabsorb the corresponding corrections into the 
definition of the other free couplings, that we choose to be of order ${\cal O}(1)$.

The fact that the lower bound in Eq.~(\ref{vRmin0}) is increasing for
$\xi\to 0$ can be understood by noticing that in the limit $\xi\to 0$ the Yukawa couplings vanish. Hence, large values of $v_R$ are needed to generate the top-quark Yukawa couplings in the small mixing $\xi \ll 1$ regime.

\begin{table} \begin{center}    
\begin{tabular}{ccccc}
\toprule
$\xi$ 
& $v^{\rm min}_R[{\rm TeV}]$
& $M^{\rm min}_{W_R}[{\rm TeV}]$
& $\bar{m}^{\rm min}[{\rm TeV}]$
& ${m}_{-}^{\rm min}[{\rm TeV}]$
\tabularnewline \midrule 
0.1
& 1581
& 169
& 554
& 526
\tabularnewline
0.2
& ~799
& ~85
& 279
& 249
\tabularnewline
0.3
& ~529
& ~56
& 185
& 155
\tabularnewline
0.5
& ~293
& ~31
& 107
& ~75
\tabularnewline
0.7
& ~164
& ~17
& ~67
& ~37
\tabularnewline
0.8
& ~110
& ~12
& ~52
& ~23
\tabularnewline
0.9
& ~~57
& ~~6
& ~35
& ~11
\tabularnewline
~\,0.95
& ~~30
& ~~3
& ~25
& ~~6
\tabularnewline\bottomrule
\end{tabular} 
\caption{ Minimum values allowed by vacuum stability and DM constraints,
for the $H_R$ VEV $v^{\rm min}_R$, the $W_R$ mass $M^{\rm min}_{W_R}$, the average messenger mass  $\bar{m}^{\rm min}$, and the lightest {\it up-down} universal messenger mass eigenvalue ${m}_{-}^{\rm min}=\bar{m}^{\rm min}\sqrt{1-\xi}$, versus some values of the mixing parameter $\xi$. Results are in unit of couplings, that is $\g=\lambda=1$.}
\label{tab2}
\end{center} 
\end{table}

\subsection{Trilinear couplings and lower bound on the
second Higgs mass}\label{ssub:trilinear}
In a first approximation the two trilinear couplings between the LH and RH
physical Higgs fields are given by
\begin{equation}
	\mu_{LRR}:=-\frac{\lambda_{LR}}{4}v_L=\frac{\lambda_{L}}{2}\frac{v^3_L}{v_R^2},\qquad \mu_{RLL}:=-\frac{\lambda_{LR}}{4}v_R=\frac{\lambda_{L}}{2}\frac{v^2_L}{v_R},
\end{equation}
where we used the expression for $\lambda_{LR}$ resulting from the
minimisation conditions in Eq.~\eqref{eq:mincon}. Clearly the process $h_L\to h_R h_R$ is then suppressed as $(v_L/v_R)^2$ with respect to the specular reaction $h_R\to h_L h_L$. Given the above estimate for the scale of $v_R$, we therefore expect the model to evade the bounds posed by the invisible decay of the SM-like Higgs boson in scenario, should this channel be kinematically allowed.

As for the potential lower bound on the $h_R$ mass, notice that under the assumption $g_L=g_R=g$, we have
$m_{W_L}^2/m_{W_R}^2=v_L^2/v_R^2$. The rough lower bound from collider searches $m_{W_R}\gtrsim1$ TeV then translates via Eq.~\eqref{eq:cond} into bound 
\beq
\sqrt{\frac{\lambda_R}{\lambda_L}}\gtrsim10^{-2}\, ,
\eeq
which, in turn, yields the following relation for the Higgs masses valid at the lowest order of the expansion in $v_L/v_r$
\beq
m_2\gtrsim 10^{-1}m_1 \approx 10 \text{ GeV}\, \, ,
\label{higgsbound}
\eeq
In the rightmost equality we have identified $m_1$ with the SM Higgs boson mass.

\subsection{The $\rho$ parameter} 
\label{sub:The rho parameter }
Considering now the tree-level contribution to the $\rho$ parameter, we can show that in the present framework $\rho=1$ up to corrections of order $ \mathcal{O}((v_L/v_R)^{2})$.
As usual we define
\bea
\rho&:=& \frac{M_{W_{L}}^2}{M^2_{Z_L} \cos^2\tw}\, ,
\label{rho}
\eea
and from the relations in Eq.~(\ref{ZL}) obtain
\bea
\cos^2\tw &=& \frac{g'^2+g^2}{2g'^2+g^2},\qquad \sin^2\tw=\frac{g'^2}{2g'^2+g^2}\, ,
\label{cosW}
\eea
and
\bea
\cos2\tw &=& \cos^2\tw-\sin^2\tw=\frac{g^2}{2g'^2+g^2}\, .
\eea
Then, by substituting Eq.~(\ref{cosW}),  Eq.~(\ref{MW}) and Eq.~(\ref{MZL}) into the 
$\rho$ parameter definition in Eq.~(\ref{rho}), we obtain $\rho=1$. 
Notice that whereas in the SM $\tan\tw^{SM}=g'/g$, here the corresponding relation is given by
\bea
\tan\tw&=&\frac{g'}{\sqrt{g'^2+g^2}}, \qquad \frac{g'}{g}=\frac{\sin\tw}{\sqrt{\cos2\tw}}\, .
\eea
The above relations are valid barring terms of order $\mathcal{O}((v_L/v_R)^{2})$, hence we expect corrections to the $\rho$ parameter to be of this order. Phenomenologically this is not a problem since these corrections are completely negligible because of the upper bound on the ratio $v_L/v_R \approx 10^{-4}$ which holds in our scenario.

\subsection{The strong CP violation problem} 
\label{sub:the strong CP violation problem}

To conclude this section, we briefly comment on the strong CP problem within the present framework. The strong CP problem arises in the SM from the necessary cancellation between the $\theta$ parameter of QCD 
and the phase of the determinant of the product of quark mass matrices $\theta_q$ \cite{Cheng:1987gp},\cite{Mohapatra:1986}.
The first appears in the Lagrangian for the QCD gluon fields from the non
trivial vacuum configurations
\begin{equation}
	\LG_{SM}\supset\theta\frac{g_S^2}{32\pi^2}\operatorname{Tr}\left(G_{\mu\nu}\tilde{G}^{\mu\nu}\right)\, ,
	\label{eq:strcp}
\end{equation}
with $g_S$ the QCD coupling, $G_{\mu\nu}$ the QCD field strength and $\tilde{G}_{\mu\nu}=\epsilon_{\mu\nu\alpha\beta}G^{\alpha\beta}$ its dual. The phase of the determinant of the product of quark mass matrices results, instead, from the $U(1)$ phase of the global flavour symmetry underlying the SM structure\footnote{For vanishing masses left-handed quark doublets and the right-handed up and down counterparts enjoy a global $U(3)^3$ symmetry. This allows at most for $U(3)$ rotations of the involved fields, which can be decomposed according to $U(3) = U(1) \times SU(3)$. The phase associated to this $U(1)$ symmetry is $\theta_q$, whereas the phases in the $SU(3)$ part are encoded in the Jarlskog invariant
\cite{Jarlskog:1985ht}.}. 
These apparently unrelated parameters are connected by the axial anomaly in a way that, consequently to a chiral rotation of the quark fields of an angle $\varphi /2$ for which the chiral $U(1)$ phase $\theta_q$ changes as $\theta_q\to\theta_q+\varphi$, the QCD $\theta$ parameter is also modified as  $\theta\to\theta+\varphi$. We can therefore perform a chiral rotation to ensure that the phase of the quark mass matrix determinant vanish and consequently identify the \textit{physical} QCD $\theta$ parameter that regulates the strong CP violation in $\bar\theta=\theta-\theta_q$.
Since a priori there is no reason to assume $\theta_q\simeq\theta$ in the SM framework, the constraint imposed by the measurements of the neutron electric dipole moment, $\bar\theta< 10^{-10}$, raises problems concerning a strong fine tuning among two uncorrelated sources of CP violation.
The usual solution to the problem is the Peccei-Quinn mechanism \cite{Peccei:1977hh}. This introduces a new $U(1)$ global symmetry
that rotates away $\bar{\theta}$ and, eventually, is spontaneously broken yielding a light pseudo Nambu-Goldstone boson in the theory: the axion \cite{Weinberg:1977ma,Wilczek:1977pj}.
Interestingly, however, the fine-tuning problem of strong CP violation can be straightforwardly solved within LR symmetric models \cite{Mohapatra:1978fy}, \cite{Diaz-Cruz:2016pmm} without invoking additional global symmetries.

Focusing for simplicity on one quark generation, after the  symmetry breaking we normally expect the presence in the the Lagrangian of terms as
\begin{equation}
	\LG \supset e^{i\theta_q} m_q \bar{q}_Lq_R + e^{-i\theta_q} m_q \bar{q}_Rq_L,
\end{equation}
and it can easily be seen that the requirement of LR symmetry, under which $q_L \leftrightarrow q_R$, forces unequivocally $\theta_q\equiv0$. As a consequence the smallness of the parameter $\bar\theta$ is here directly related to the smallness of the orignal QCD $\theta$ parameter and the ansatz $\bar\theta=\theta=0$ does not raise any fine-tuning problem. 
This general argument shows that the problem of strong CP violation arises from the violation of the parity symmetry. Once the latter is restored, as for instance in LR symmetric extension of the SM, the measurement of CP violations in the strong sector directly relate to the magnitude of the term in Eq.~\eqref{eq:strcp}. 

To see this aspect in more detail, consider the effective $\bar{\theta}$ parameter in the present framework
\bea
\bar{\theta}=\theta+\arg[{\rm det}\left(M_{\U} M_{\D}\right)]\, ,
\label{theta}
\eea
where $M_{\U}$ and $M_{\D}$ are the quark mass matrices for up- and down-quark sectors respectively\footnote{Notice that the weak CP violations resides in the Jarlskog invariant $J$
\cite{Jarlskog:1985ht}, which is proportional to the imaginary part of the following commutator
\bea
J\sim {\rm Im}\left[\det([M_{\U} M_{\U}^{\dag},M_{\D} M_{\D}^{\dag}])\right]\, .
\eea
Therefore, if $\arg[\det\left(M_u M_d\right)]=0$, one can still have $J\neq 0$, allowing for a non-vanishing contributions to the weak CP violation.}.

By diagonalising the quark mass matrices
\bea
{\rm diag}[M_{\U,\D}]=V^{{\U},{\D}~\dag}_{L}\cdot M_{\U,\D}\cdot V^{\U,\D}_{R}\, ,
\eea
we can see that
\bea
\arg[\det\left(M_{\U} M_{\D}\right)]=(\alpha^{\U}_R-\alpha^{\U}_L)+(\alpha^{\D}_R-\alpha^{\D}_L)\, ,
\eea
where  $V^{{\U},{\D}~\dag}_{L,R}$ are unitary matrices and $\alpha^{\U,\D}_{L,R}=\arg\det(V_{L,R}^{\U,\D})$. Therefore, the strong CP problem is straightforwardly solved if the following conditions are simultaneously satisfied \cite{Diaz-Cruz:2016pmm}
\bea
\alpha^{\U}_{L}&=&\alpha^{\U}_{R}\, , ~~~~~\alpha^{\D}_{L}\,=\,\alpha^{\D}_{R}\, .
\label{alphaCP}
\eea

In order to show that this is indeed the case in our setup, we first need to generalize the interacting Lagrangian in Eq.~(\ref{LagMS}) to the flavour mixing effects encapsulated in the CKM matrix. As explained in \cite{Gabrielli:2016cut}, this can be done by extending the Lagrangian of Eq.~(\ref{LagMS}) to
\bea
{\cal L}^I_{MS} &\supset&
\hat{g}_L\left( \sum_{i,j=1}^{N}\left[\bar{q}^i_L (X_L^U)_{ij} Q_R^{\U_j}\right] \hat{S}^{\U_j}_{L} +
\sum_{i,j=1}^{N}\left[\bar{q}^i_L  (X_L^D)_{ij} Q_R^{\D_j}\right] \hat{S}^{D_j}_{L}\right)
\nonumber\\
&+&
\hat{g}_R\left( \sum_{i,j=1}^{N}\left[\bar{q}^i_R (X_R^U)_{ij} Q_L^{\U_j}\right] \hat{S}^{\U_j}_{R} +
\sum_{i,j=1}^{N}\left[\bar{q}^i_R  (X_L^D)_{ij} Q_L^{\D_j}\right] \hat{S}^{D_j}_{R}\right)
\label{LagMS-CKM}
\eea 
where $X_{L,R}^{U,D}$ are generic matrices. Notice
that the $U(1)_F$ gauge invariance and non-universality of $U(1)_F$ charges  
require the family index that label dark fermions and scalar messengers to be the same. Then, in the weak-current basis of quark fields, 
the Yukawa couplings generated radiatively follow the pattern
\bea
Y^{U,D}_{ij}\sim \left( X^{U,D~\dag}_L\cdot \hat{Y}^{U,D}\cdot X^{U,D}_R\right)_{ij}\, ,
\label{Ygeneral}
\eea
where the $\,\cdot \,$ symbol stands for a matrix product,  
$\hat{Y}^{U,D}={\rm diag}[Y^{U,D}_1,Y^{U,D}_2,Y^{U,D}_3]$, and 
$Y_i^{U,D}$ ($i=1,2,3$) are the Yukawa couplings of the diagonal case in Eq.~(\ref{Yukexact}) for the {\it up} and {\it down} sectors
respectively. The observed structure of the CKM matrix forces the $X_{L,R}^{U,D}$ to be proportional to the unity matrix ${\bf 1}$ in the family space barring suppressed off-diagonal entries \cite{Gabrielli:2016cut}:
\bea
X_{L,R}^{U,D}\sim {\bf 1} + \Delta_{L,R}^{U,D}\, ,
\label{X}
\eea
being $|\Delta_{L,R}^{U,D}|_{ij}\ll 1$. We remark that Eq.~(\ref{X}) clearly shows that the matrices $X_{L,R}^{U,D}$ are in general not unitary and, in particular, do not result from a basis rotation\footnote{For the mechanism that can generate the required hierarchy in the off-diagonal terms  $|\Delta_{L,R}^{U,D}|$ see \cite{Gabrielli:2016cut}.}. 

Imposing the LR symmetry on the Lagrangian in Eq.~(\ref{LagMS-CKM}) then yields 
\bea
\hat{g}_L&=\hat{g}_R \, , ~~~~X^{U,D}_L\, =\, X^{U,D}_R\,\equiv \, X^{U,D}\, .
\eea
and consequently, owing to the LR symmetry, 
the Yukawa matrices in Eq.~(\ref{Ygeneral}),  
as well as the corresponding quark matrices $M_{\U,\D}$, are necessarily 
hermitian matrices: 
\bea
Y^{U,D}_{ij}\sim \left( X^{U,D~\dag}\cdot \hat{Y}^{U,D}\cdot X^{U,D}\right)_{ij}\,  .
\label{YLR}
\eea
The diagonalizations of $Y^{\U,\D}$, or equivalently of the $M_{\U,\D}$ mass matrices, then involve only two unitary hermitian matrices $V^{\U}\equiv V^{\U}_L=V^{\U}_R$ and $V^{\D}\equiv V^{\D}_L=V^{\D}_R$, that respectively operate on the up- and down-quark fields respectively, and the condition in Eq.~(\ref{alphaCP}) is therefore trivially satisfied.

\section{Discussion and Conclusions} 
\label{sec:Conclusions}

Left-right symmetric models are amongst the very first extensions of the SM which were proposed and, therefore, have been thoroughly scrutinised in time.
The main motivation supporting the $SU(2)_L \times SU(2)_R \times U(1)_{B-L}$ gauge group lies in the magnitude of parity violation observed in weak interactions, which in the left-right symmetric framework is explained by 
the largeness of the right-handed symmetry breaking scale. Other virtues of the left-right symmetric framework include, for instance, a possible solution to the strong CP-problem and the explanation of the SM hypercharge quantum numbers in terms of baryonic and leptonic charges.

In this work we yet proposed a new reason to adopt the left-right symmetry that, to our knowledge, has not been considered before.
Under the assumption that the SM Yukawa couplings are effective low-energy quantities, we investigated a left-right symmetric framework characterised by the simplest scalar content able to explain the large hierarchy between the left and the right symmetry breaking scales. Our construction presents one left-handed and one right-handed Higgs doublets, and insures the radiative origin of Yukawa couplings by forbidding the presence of Higgs fields in the bidoublet representation of the gauge group. 

In order to assess the viability of the proposed setup, in this paper we investigated the properties of the gauge and Higgs bosons of the theory, assessing also their impact on the electroweak precision observables. In this regard, we find that the considered electroweak observables receive corrections of order $\mathcal{O}(v_L^2/v_R^2)$, being $v_L$ and $v_R$ the vacuum expectation values of the corresponding Higgs doublets. Given that the current collider bounds on additional gauge bosons force here the right-handed symmetry breaking scale to be large, we find that the mentioned constraints are negligible. Interestingly, the hierarchy of the scales $v_L\ll v_R,$ is imputable in our frame to the smallness of the portal coupling $\lambda_{LR}$ between the two sectors. We also find that in spite of the lower bound on $v_R$, small mass values for the right-handed Higgs doublet are \textit{not} forbidden owing to a suppressed scalar self-coupling. Given that the $SU(2)_R$ gauge bosons are expected to be heavy in this scenario, this feature potentially allows for tests of the framework at the LHC and future collider experiments.

Clearly, in order to comply with the experimental results, our framework must also reproduce the observed SM Yukawa couplings. For the chosen setup of the Higgs sector, these quantities are necessarily generated at the loop level after the spontaneous symmetry breaking of the symmetry in the right-handed sector of the theory. A distinctive feature of our scenarion is that the necessary chiral symmetry breaking that sources the low-energy generation of Yukawa couplings is provided by a hidden sector of the theory. In the example we discussed, chiral symmetry breaking arises from the masses of dark fermions, the lightest of which plays the role of Dark Matter candidate. The chiral symmetry breaking is then transferred to the Standard Model by a set of scalar messangers at the loop level and, owing to non-perturbative dynamics in the hidden sector, this framework can give rise to the observed spread of Standard Model fermion masses in a natural way.

In spite of the adopted setup for the Higgs sector, we find in our investigation that our scenario straightforwardly solves the strong CP problem. The underlying left-right symmetry enforces in fact the cancellation of phases that would otherwise result in a sizeable QCD $\theta$ parameter through the chiral anomaly. As typical in left-right models, the hypercharge gauge group can be furthermore identified here with $B-L$. 

It is instructive to compare our results with the attempts of radiative Yukawa generation made within supersymmetric models, in which the loop-level Yukawa couplings are induced by soft supersymmetry breaking parameters as the A-terms and the squark masses. Because the naturalness of supersymmetric models requires the latter to be less than ${\cal O}(1)$~TeV, only the Yukawa couplings of the first two Standard Model generation can be radiatively generated. The top and bottom quarks must instead obtain their masses at tree level. On the contrary, in our case, the Yukawa couplings of all the Standard Model fermions can be generated radiatively due to the large right-handed breaking scale. Notice also that no tuning of massive parameters is needed in our model to explain the smallness of the electroweak scale compared with the right-handed breaking scale.
To conclude, we believe that the proposed left-right symmetric framework characterised by the simplest viable Higgs sector provides a fruitful working ground to address the SM flavour issues in the phenomenologically testable way.

\section*{Appendix:  examples of hidden sector dynamics}

In this Appendix we sketch potential hidden sectors in the framework of \cite{Gabrielli:2013jka} that result in hierarchical effective Yukawa couplings for the SM fermions. 

Our previous discussion made clear that the masses of the dark fermions in the hidden sector supply the chirality violation necessary for the SM Yukawa coupling generation. As a consequence, provided that the messenger sector has a flovour-blind structure, the Yukawa couplings emerging from the hidden sector dynamics are necessarily proportional to the dark fermion masses. The hierarchy observed in the former is then imputable to the spread of the latter, and the problem of the SM flavour hierarchy is solved in presence of a mechanism that gives rise to the required hierarchical dark fermion mass spectrum.

\subsubsection*{A) Hierarchy from higher derivatives} 
\label{ssub:A weakly interacting theory}

In \cite{Gabrielli:2013jka}, the mechanism behind the dark fermions mass spectrum generation is based on non-perturbative dynamics inspired by the Nambu-Jona-Lasinio approach. More in detail, the hidden sector comprises dark fermions characterised by non-universal charges associated to a new $U(1)_F$ local gauge symmetry, under which the SM fields are all singlets. The Lagrangian for the dark fermions and the dark photon of the $U(1)_F$ gauge group is given by, \cite{Gabrielli:2013jka} 
\bea
{\cal L}_{DS}&=& 
i\sum_i \left( 
\bar{Q}^{U_i}{\cal D}_{\mu}\gamma^{\mu} Q^{U_i}+\bar{Q}^{D_i}{\cal D}_{\mu}\gamma^{\mu} Q^{D_i}\right)
\nonumber \\
&-&
\frac{1}{4} F_{\mu\nu} F^{\mu\nu} + \frac{1}{2\Lambda^2}  \partial^{\mu} F_{\mu\alpha} \partial_{\nu} F^{\nu\alpha},
\label{LagDS}
\eea
where $Q^{U_i}$, $Q^{D_i}$ are the dark fermion fields, partners in the hidden sector of the SM up ($U_i$) and down ($D_i$) quarks. We indicated with ${\cal D}_{\mu}=\partial_{\mu}+i g \hat{Q} \bar{A}_{\mu}$ the covariant derivative involving the dark-photon $\bar{A}_{\mu}$ with associated field strength tensor $F_{\mu\alpha}$, whereas $\hat{Q}$ is the charge operator acting on the dark fermion fields. The Lagrangian ${\cal L}_{DS}$ can be extended to include also the SM leptonic sector in a straightforward way.

Differently from the traditional QED Lagrangian, ${\cal L}_{DS}$ presents a pure $U(1)_F$ gauge term which contains higher-derivatives, the so-called Lee-Wick term, hence the scale $\Lambda$ can be interpreted as the mass of the associated ghost field. As shown in \cite{Gabrielli:2007cp}, this term is crucial for triggering the required chiral symmetry breaking in a weak coupling regime. In fact, by solving the mass-gap equation in the Nambu-Jona-Lasinio  \cite{Nambu:1961tp} approach, the presence of such term ensures the existence of a non-trivial solution for the dark fermion mass spectrum on the true vacuum, given by \cite{Gabrielli:2007cp}
\bea
M_{Q_i}=\Lambda \exp\left\{-\frac{2\pi}{3\,\bar{\alpha}(\Lambda) q_i^2} +\frac{1}{4}\right\}\, ,
\label{mgap2}
\eea
where $q_i$ is the $U(1)_F$ charge of the considered dark fermion and $\bar{\alpha}(\Lambda)$ is the corresponding fine structure constant evaluated at the  $\Lambda$ energy scale associated to the ghost field. The solution is manifestly non-perturbative, as testified by the dependence of $M_{Q_i}$ on $\bar{\alpha}$, and can give rise to an exponentially spread dark fermion mass spectrum provided that the latter have different $U(1)_F$ charges.

By assuming that the messenger sector is flavour-blind, the only source of flavour violation comes from the $U(1)_F$ charges. As a consequence, it is possible to relate the ratios of SM quark masses to the $\bar{\alpha}(\Lambda)$ and $U(1)_F$ charges of the corresponding dark fermions. In particular, if we define $\bar{\alpha}(\Lambda)$ by normalizing to 1 the largest $U(1)_F$ charge, associated to the dark-fermion partner of the top quark, according to Eqs.(\ref{MQi}) and (\ref{mgap2}), we obtain the following
mass sum rule
\bea
\bar{\alpha}(\Lambda)^{-1} &\simeq &\frac{3}{2\pi} \frac{q_{Q_i}^2}{1-q_{\Q_i}^2}
\log\frac{m_t}{m_i}\, ,
\label{sumrules1}
\eea
where $m_t$ is the top-quark mass,  $m_i$ is the mass of a generic SM quark $Q_i$ and $q_{\Q_i}$ is the $U(1)_F$ charge of the dark fermion partner of $Q_i$. In deriving the above relation we have used the property that the loop function $f_1(x_i,\xi)$, as defined in Eq.(\ref{f1}) with $x_i=M_{Q_i}^2/\bar{m}^2$, depends weakly on the dark-fermion mass $M_{Q_i}$,  hence it can be approximated as a constant and ratios of masses cancel out. A generalization of Eq.(\ref{f1}) to lepton masses is straightforward.

By fixing  the ratio of two $q_{\Q_i}$ charges in Eq.(\ref{sumrules1}), we can then predict $\bar{\alpha}(\Lambda)$, as well as all the remaining $U(1)_F$ charges. For instance, by setting the ratio $q_{\D_3}/q_{\U_3}=0.9$ and by normalizing $q_{\U_3}=1$,
from Eq.(\ref{sumrules1}) we obtains $\bar{\alpha}(\Lambda) \simeq  0.14$.
Reproducing larger splitting for the charges would require larger values of  $\bar{\alpha}$.
Once $\bar{\alpha}$ is known, we can arrange the remaining $U(1)_F$ charges of the dark fermion in a way that the masses of the corresponding SM quarks and lepton match the measured values \footnote{ Notice that in Eq(\ref{sumrules1}) one should actually use the running quark masses evaluated at the messenger mass scale, although for simplicity we will approximate them here with the corresponding quark mass poles.}. As an example, using $\bar{\alpha}(\Lambda)=0.14$ obtained for
$q_{\U_3}=1$ and $q_{\D_3}=0.9$, we report in table~\ref{tab:chex} the $U(1)_F$ charges yielding the central values of heavy quarks (top, bottom, and charm) pole masses, as well as the central values of the remaining quark masses, as computed in the $\overline{\rm MS}$ scheme evaluated at a renormalization scale $\mu=2$ GeV \cite{Olive:2016xmw}.

Repeating the exercise for the SM leptons, we obtain the results in Table~\ref{tab:chexl}. In this calculation we set $q_{\E_3}=1$ and considered a reference neutrino hierarchy given by $m_{N_1}=1$~eV, $m_{N_2}=10^{-3}$~eV, and $m_{N_3}=10^{-6}$~eV. Notice that for both the cases of quarks and leptons, the value of the charges are such that the theory remains well within the boundaries of a weakly-coupled regime.

\begin{table}[htb!]
	\label{tab:chex}
	\centering
	\begin{tabular}{c|c|c|c|c}
		\toprule
		$\psi$: & $U_2$ & $D_2$ & $U_1$ & $D_1$\\
		\midrule
		$q_\psi$ & 0.88 & 0.82 &0.77& 0.76\\
		\bottomrule
	\end{tabular}
	\caption{Values of the $U(1)_F$ dark fermion charges resulting in the observed SM quark mass hierarchy for $\bar{\alpha}(\Lambda)=0.14$, $q_{\U_3}=1$ and $q_{\D_3}=0.9$. The first line indicates the associated SM quark field and the second line the $U(1)_F$ charge of the corresponding dark fermion partner.}
\end{table}

\begin{table}[htb!]
	\label{tab:chexl}
	\centering
	\begin{tabular}{c|c|c|c|c|c}
		\toprule
		$\psi$: & $E_2$ & $E_1$ & $N_3$ & $N_2$ & $N_1$\\
		\midrule
		$q_\psi$ & 0.92 & 0.81 &0.65& 0.59& 0.55\\
		\bottomrule
	\end{tabular}
	\caption{Values of the $U(1)_F$ dark fermion charges resulting in the observed SM lepton mass hierarchy, obtained with the normalization choice $q_{\E_3}=1$ and a reference neutrino hierarchy given by $m_{N_1}=1$~eV, $m_{N_2}=10^{-3}$~eV, and $m_{N_3}=10^{-6}$~eV. The first line indicates the associated SM lepton field and the second line the $U(1)_F$ charge of the corresponding dark fermion partner.}
\end{table}

\subsubsection*{B) The possible role of strongly coupled dynamics } 
\label{sub:A model with strongly interacting dark sector}

The hidden sector proposed in \cite{Gabrielli:2013jka} can embed another possible mechanism to achieve the required spread of the dark fermion masses. Consider the following Lagrangian  

\bea
{\cal L}_{DS}^\prime&=& 
i
\bar\psi{\cal D}_{\mu}\gamma^{\mu} \psi
-
\frac{1}{4} F_{\mu\nu} F^{\mu\nu},
\label{LagDSp}
\eea
where for simplicity we focused on one massless dark fermion field $\psi$
and here ${\cal D}_{\mu}=\partial_{\mu}+i q_\psi \bar{A}_{\mu}$ the covariant derivative involving the dark-photon $\bar{A}_{\mu}$. As before $F_{\mu\alpha}$ is the relative field strength tensor.
Although the above Lagrangian matches the one of ordinary QED for massless fermions, we can employ non perturbative effect arising in the strongly coupled regime of the theory to generate a dynamical mass for the dark fermion. More in detail, for supercritical values of the dark fine-structure constant $\alpha = q_\psi^2/4\pi  > \alpha_c = \pi/3$, solving the gap equation for the dark fermion mass $m_\psi$ yields a non-trivial solution, which exhibits an exponential scaling behavior (Miransky scaling) \cite{Fomin:1978rk, Miransky:1984ef}
\begin{equation}
	m_\psi \approx 4\Lambda_c\,e^{-\dfrac{\pi\alpha_c}{\sqrt{\alpha-\alpha_c}}}.
\end{equation}
Notice that differently from the case of QED with higher derivative terms discussed before, the scale $\Lambda_c$ emerges spontaneously from the theory and is associated to the energy scale at which the present theory is dominated by strongly coupled dynamics.

Extending the above model to the full particle content of the hidden sector, it is then conceivable that the required dark fermion mass spectrum could be achieved by arranging the charges of the latter maintaining the theory in a supercritical regime. 

\section*{Acknowledgements} 
\label{sec:Acknowledgements}
The authors thank Kristjan Kannike for useful discussions.
EG would like to thank the CERN Theory Division for the kind hospitality during the preparation of this work. This work was supported by the Estonian Research Council with the grants  IUT23-6, PUTJD110 and by EU through the ERDF CoE program.



\begin{thebibliography}{99}

\bibitem{Aad:2012tfa}
  G.~Aad {\it et al.}  [ATLAS Collaboration],
  Phys.\ Lett.\ B {\bf 716}, 1 (2012)
  [arXiv:1207.7214 [hep-ex]];
  S.~Chatrchyan {\it et al.}  [CMS Collaboration],
  Phys.\ Lett.\ B {\bf 716}, 30 (2012)
  [arXiv:1207.7235 [hep-ex]].


\bibitem{Englert:1964et}
  F.~Englert and R.~Brout,
  Phys.\ Rev.\ Lett.\  {\bf 13}, 321 (1964);
  P.~W.~Higgs,
  Phys.\ Lett.\  {\bf 12}, 132 (1964);
  P.~W.~Higgs,
  Phys.\ Rev.\ Lett.\  {\bf 13}, 508 (1964);
  G.~S.~Guralnik, C.~R.~Hagen and T.~W.~B.~Kibble,
  Phys.\ Rev.\ Lett.\  {\bf 13}, 585 (1964).

\bibitem{Froggatt:1978nt} 
  C.~D.~Froggatt and H.~B.~Nielsen,
  Nucl.\ Phys.\ B {\bf 147}, 277 (1979).
  
\bibitem{Gabrielli:2013jka} 
  E.~Gabrielli and M.~Raidal,
  Phys.\ Rev.\ D {\bf 89}, 015008 (2014)
  [arXiv:1310.1090 [hep-ph]].

\bibitem{Davidson:1987mh} 
  A.~Davidson and K.~C.~Wali,
  Phys.\ Rev.\ Lett.\  {\bf 59}, 393 (1987).
  doi:10.1103/PhysRevLett.59.393
  
\bibitem{Brahmachari:2003wv} 
  B.~Brahmachari, E.~Ma and U.~Sarkar,
  Phys.\ Rev.\ Lett.\  {\bf 91}, 011801 (2003)
  doi:10.1103/PhysRevLett.91.011801
  [hep-ph/0301041].

\bibitem{Ma:2014rua} 
  E.~Ma,
Phys. Rev. Lett.,112,091801;
  S.~Fraser and E.~Ma,
  Europhys.\ Lett.\  {\bf 108}, 1002 (2014)
  doi:10.1209/0295-5075/108/11002
  [arXiv:1402.6415 [hep-ph]].

\bibitem{preparation} 
E. Gabrielli, M. Raidal, and L. Marzola, in preparation.


\bibitem{Mohapatra:1974gc}
  R.~N.~Mohapatra and J.~C.~Pati,
  Phys.\ Rev.\ D {\bf 11} (1975) 2558.
  doi:10.1103/PhysRevD.11.2558

\bibitem{Mohapatra:1974hk} 
  R.~N.~Mohapatra and J.~C.~Pati,
  Phys.\ Rev.\ D {\bf 11}, 566 (1975).
  doi:10.1103/PhysRevD.11.566

\bibitem{Senjanovic:1975rk} 
  G.~Senjanovic and R.~N.~Mohapatra,
  Phys.\ Rev.\ D {\bf 12}, 1502 (1975).
  doi:10.1103/PhysRevD.12.1502


\bibitem{Mohapatra:1977mj} 
  R.~N.~Mohapatra, F.~E.~Paige and D.~P.~Sidhu,
  Phys.\ Rev.\ D {\bf 17}, 2462 (1978).
  doi:10.1103/PhysRevD.17.2462

\bibitem{Senjanovic:1978ev} 
  G.~Senjanovic,
  Nucl.\ Phys.\ B {\bf 153}, 334 (1979).
  doi:10.1016/0550-3213(79)90604-7

\bibitem{Duka:1999uc} 
  P.~Duka, J.~Gluza and M.~Zralek,
  Annals Phys.\  {\bf 280}, 336 (2000)
  doi:10.1006/aphy.1999.5988
  [hep-ph/9910279].


\bibitem{Tello:2010am} 
  V.~Tello, M.~Nemevsek, F.~Nesti, G.~Senjanovic and F.~Vissani,
  Phys.\ Rev.\ Lett.\  {\bf 106}, 151801 (2011)
  doi:10.1103/PhysRevLett.106.151801
  [arXiv:1011.3522 [hep-ph]].

\bibitem{Maiezza:2010ic} 
  A.~Maiezza, M.~Nemevsek, F.~Nesti and G.~Senjanovic,
  Phys.\ Rev.\ D {\bf 82}, 055022 (2010)
  doi:10.1103/PhysRevD.82.055022
  [arXiv:1005.5160 [hep-ph]].

\bibitem{Mohapatra:1978fy} 
  R.~N.~Mohapatra and G.~Senjanovic,
  Phys.\ Lett.\ B {\bf 79}, 283 (1978).
  doi:10.1016/0370-2693(78)90243-5;
  R.~N.~Mohapatra, A.~Rasin and G.~Senjanovic,
  Phys.\ Rev.\ Lett.\  {\bf 79} (1997) 4744
  doi:10.1103/PhysRevLett.79.4744
  [hep-ph/9707281].

\bibitem{Diaz-Cruz:2016pmm} 
  J.~L.~Díaz-Cruz, W.~G.~Hollik and U.~J.~Saldaña-Salazar,
  arXiv:1605.03860 [hep-ph].


\bibitem{Peccei:1977hh} 
  R.~D.~Peccei and H.~R.~Quinn,
  Phys.\ Rev.\ Lett.\  {\bf 38}, 1440 (1977).
  doi:10.1103/PhysRevLett.38.1440

\bibitem{Weinberg:1977ma} 
  S.~Weinberg,
  Phys.\ Rev.\ Lett.\  {\bf 40}, 223 (1978).
  doi:10.1103/PhysRevLett.40.223

\bibitem{Wilczek:1977pj} 
  F.~Wilczek,
  Phys.\ Rev.\ Lett.\  {\bf 40}, 279 (1978).
  doi:10.1103/PhysRevLett.40.279


\bibitem{Branco:2011iw} 
  G.~C.~Branco, P.~M.~Ferreira, L.~Lavoura, M.~N.~Rebelo, M.~Sher and J.~P.~Silva,
  Phys.\ Rept.\  {\bf 516}, 1 (2012)
  doi:10.1016/j.physrep.2012.02.002
  [arXiv:1106.0034 [hep-ph]].


\bibitem{Gabrielli:2007cp} 
  E.~Gabrielli,
  Phys.\ Rev.\ D {\bf 77}, 055020 (2008)
  [arXiv:0712.2208 [hep-ph]].


\bibitem{Nambu:1961tp} 
  Y.~Nambu and G.~Jona-Lasinio,
  Phys.\ Rev.\  {\bf 122}, 345 (1961);
  Y.~Nambu and G.~Jona-Lasinio,
  Phys.\ Rev.\  {\bf 124}, 246 (1961).

\bibitem{Lee:1971ix} 
  T.~D.~Lee and G.~C.~Wick,
  Phys.\ Rev.\ D {\bf 3}, 1046 (1971).
  T.~D.~Lee and G.~C.~Wick,
  Phys.\ Rev.\ D {\bf 2}, 1033 (1970).

\bibitem{Grinstein:2007mp} 
  B.~Grinstein, D.~O'Connell and M.~B.~Wise,
  Phys.\ Rev.\ D {\bf 77}, 025012 (2008)
  [arXiv:0704.1845 [hep-ph]].
  
\bibitem{Gabrielli:2016cut} 
  E.~Gabrielli, B.~Mele, M.~Raidal and E.~Venturini,
  arXiv:1607.05928 [hep-ph].

\bibitem{Gabrielli:2014oya} 
  E.~Gabrielli, M.~Heikinheimo, B.~Mele and M.~Raidal,
  Phys.\ Rev.\ D {\bf 90}, 055032 (2014)
  [arXiv:1405.5196 [hep-ph]];
  S.~Biswas, E.~Gabrielli, M.~Heikinheimo and B.~Mele,
  Phys.\ Rev.\ D {\bf 93}, no. 9, 093011 (2016)
  doi:10.1103/PhysRevD.93.093011
  [arXiv:1603.01377 [hep-ph]];
  S.~Biswas, E.~Gabrielli, M.~Heikinheimo and B.~Mele,
  JHEP {\bf 1506}, 102 (2015)
  doi:10.1007/JHEP06(2015)102
  [arXiv:1503.05836 [hep-ph]].



\bibitem{Cheng:1987gp} 
  H.~Y.~Cheng,
  Phys.\ Rept.\  {\bf 158}, 1 (1988).
  doi:10.1016/0370-1573(88)90135-4.
\bibitem{Mohapatra:1986} 
  R. N. Mohapatra, Unification and Supersymmetry. The Frontiers of Quark-Lepton
  Physics. Springer, Berlin, 1986.

\bibitem{Jarlskog:1985ht} 
  C.~Jarlskog,
  Phys.\ Rev.\ Lett.\  {\bf 55}, 1039 (1985).
  doi:10.1103/PhysRevLett.55.1039

\bibitem{Olive:2016xmw} 
  C.~Patrignani {\it et al.} [Particle Data Group],
  Chin.\ Phys.\ C {\bf 40}, no. 10, 100001 (2016).
  doi:10.1088/1674-1137/40/10/100001

\bibitem{Fomin:1978rk} 
  P.~I.~Fomin, V.~P.~Gusynin and V.~A.~Miransky,
  Phys.\ Lett.\  {\bf 78B}, 136 (1978).
  doi:10.1016/0370-2693(78)90366-0

\bibitem{Miransky:1984ef} 
  V.~A.~Miransky,
  Nuovo Cim.\ A {\bf 90}, 149 (1985).
  doi:10.1007/BF02724229
\end{thebibliography}
\end{document}